\documentclass[useAMS,usenatbib]{mn2e}
\usepackage{amsmath,amstext,amsgen,amsbsy,amsopn,amsfonts,theorem}
\usepackage{natbib}
\pdfoutput=1

\usepackage{graphicx}
\usepackage{xspace}
\usepackage{amsmath}
\usepackage{hyperref}
\usepackage{framed}
\usepackage{txfonts}
\usepackage{epstopdf}
\usepackage{gensymb}

\usepackage[normalem]{ulem}

\makeatletter
\def\mr@ignsp#1 {\ifx\:#1\@empty\else #1\expandafter\mr@ignsp\fi}%
\newcommand{\multiref}[1]{\begingroup
\xdef\mr@no@sparg{\expandafter\mr@ignsp#1 \: }%
\def\mr@comma{}%
\@for\mr@refs:=\mr@no@sparg\do{\mr@comma\def\mr@comma{,}\ref{\mr@refs}}%
\endgroup}
\makeatother

\newcommand{\hypref}[2]{\ifx\href\asklfhas #2\else\href{#1}{#2}\fi}
\newcommand{\Secref}[1]{Section~\multiref{#1}}
\newcommand{\secref}[1]{Sec.~\multiref{#1}}
\newcommand{\Appref}[1]{Appendix~\multiref{#1}}

\newcommand{\Figref}[1]{Figure~\multiref{#1}}
\newcommand{\figref}[1]{Fig.~\multiref{#1}}
\renewcommand{\eqref}[1]{(\multiref{#1})}

\newcommand{\eq}[1]{\begin{align}#1\end{align}}

\newcommand{\nln}{\nonumber\\}

\title[Orbits and streams in aspherical potentials]{Stray, swing and scatter: angular momentum evolution of
  orbits and streams in aspherical potentials}

\author[D. Erkal, J. L. Sanders \& V. Belokurov]
  {Denis Erkal$^1$\thanks{derkal@ast.cam.ac.uk}, Jason L. Sanders$^1$\thanks{jls@ast.cam.ac.uk} \& Vasily Belokurov$^1$\thanks{vasily@ast.cam.ac.uk} \\
  $^1$Institute of Astronomy, Madingley Road, Cambridge, CB3 0HA, UK}

\begin{document}

\label{firstpage}

\maketitle

\begin{abstract}
In aspherical potentials orbital planes continuously evolve. The
gravitational torques impel the angular momentum vector to precess,
that is to slowly stray around the symmetry axis, and nutate,
i.e. swing up and down periodically in the perpendicular
direction. This familiar orbital pole motion - if detected and
measured - can reveal the shape of the underlying gravitational
potential, the quantity only crudely gauged in the Galaxy so far. Here
we demonstrate that the debris poles of stellar tidal streams show a
very similar straying and swinging behavior, and give analytic
expressions to link the amplitude and the frequency of the pole
evolution to the flattening of the dark matter distribution. While these results are derived for near-circular orbits, we show they are also valid for eccentric orbits. Most
importantly, we explain how the differential orbital plane precession
leads to the broadening of the stream and show that streams on polar
orbits ought to scatter faster. We provide expressions for the stream
width evolution as a function of the axisymmetric potential flattening and the
angle from the symmetry plane and prove that our models are in good
agreement with streams produced in N-body simulations. Interestingly,
the same intuition applies to streams whose progenitors are on short
or long-axis loops in a triaxial potential. Finally, we present a
compilation of the Galactic cold stream data, and discuss how the simple
picture developed here, along with stream modelling, can be used to constrain the
symmetry axes and flattening of the Milky Way.
\end{abstract}


\section{Introduction}

The stretched appearance of stellar streams is the result of the
accumulated differential rotation shear that stars with slightly
different energies acquire after they have left their disrupting
progenitor. The tendency of tidal tails to extend primarily in one
dimension was pointed out by \citet{johnston_1998} who presented an
insightful picture in which the azimuthal debris diffusion was the
result of the perturbation of the angular frequency of stripped
stars. Additionally, the influence of the disk in the host potential
was singled out as the source of the progenitor's orbital plane
precession leading to pronounced stream
broadening. \citet{helmi_white_1999} generalised this idea by invoking
the conservation of the phase-space density stipulated by Liouville's
theorem, thus building an elegant stream evolution model based on the
action-angle formalism. Irrespective of the framework chosen, the two
primary stream properties, its length and its width are thus shown to
be controlled by the progenitor's energy reservoir (or its frequency
range) and the details of the host's potential.

Taking forward the ideas introduced above, rapid advances have been
made in the last five years to describe the stream centroid behaviour
\citep[see
  e.g.][]{eyrebinney11,varghese2011,sanders2014,price_whelan_et_al_rewinder,bovy2014,gibbons_et_al_2014,bowden_gd1,fardal2015}. In addition, progress has been made in understanding how the stream expands within the orbital plane due to differential apsidal precession \citep{johnston_et_al_2001,amorisco_2014,hendel_johnston_2016}. This adds a dimension to the picture of stream growth and if the growth within the plane is sufficiently rapid, the debris will in fact resemble a shell instead of a stream. However, the evolution of the stream debris in the third dimension, perpendicular to the orbital plane, has not been directly explored. Here, we strive to fill this gap with an intuitive model of stream fanning in aspherical potentials. There are
of course, several mechanisms that can lead to the broadening of the
stellar debris perpendicular to the stream plane. For example, repeated weak
interactions with dark matter subhaloes can cause streams to become
thicker \citep[see e.g.][]{ibata_et_al_2002,ngan2016}. Additionally,
streams exploring chaotic regions of the host potential will find it
difficult to maintain coherence \citep[see
  e.g.][]{price_whelan_et_al_chaos}. Having said that, here we
concentrate on what we believe is the primary cause of stream swelling
in a large range of potentials - the differential orbital plane precession.

It is clear that streams have to have non-zero widths as the stripped
stars spray out of the Lagrange points on orbits misaligned with
respect to that of the progenitor. In spherically symmetric hosts,
where the angular momentum is conserved, the average stream width will
remain unchanged, even though stars in the stream will oscillate
around the progenitor's orbit. However, in flattened potentials, each
star's angular momentum vector will start to wander around the
symmetry axis. This continuous precession is complemented by the
pole's periodic swinging in the perpendicular direction,
i.e. nutation. While the nutation adds to the overall width of the
stream, its contribution does not increase with time. The differential
orbital plane precession, on the contrary, will induce the stellar debris
to disperse more and more in the direction perpendicular to the stream plane, steadily
broadening the stream and diluting its surface brightness.

To describe the connection between the potential flattening and the
stream width, we first consider the evolution of the angular momentum
of an individual orbit. Namely, for an arbitrarily oriented orbit in a
potential with little flattening, we write down analytic expressions
for the precession and nutation frequencies as well as the nutation
amplitude. Note that the question of the orbital plane precession has
been addressed many a time in other astrophysical situations. These
include, for example, the classical question of the nodal precession
of the Moon \citep[see e.g.][]{brown1896,MurrayDermott}, and exoplanets in general,
as well as warping of accretion \citep[e.g.][]{larwood1996} and
galactic disks \citep[see e.g.][]{steiman_durisen_1990}. Most
importantly, as we show below, the debris planes of stellar streams
exhibit very similar behaviour, i.e. the stream pole precesses and
nutates, and thus it should be possible to infer the shape of the
underlying potential given a set of accurate measurements of the
stream's structural properties. We also note that analytic expressions similar to the ones below
for the precession rates of streams have been written before in
\cite{ibata_et_al_2001} \citep[using equations
  from][]{steiman_durisen_1990}. While there is an agreement - at
first order - between the precession rate formulae presented here and
those in \cite{ibata_et_al_2001}, the latter are only valid for a
special case of a flattened logarithmic potential.

Currently, only one stream in the Milky Way has had its debris plane
precession measured, namely that from the Sagittarius dwarf galaxy. \cite{ibata_et_al_2001} used the lack of plane precession measured from Carbon stars in the stream to put limits on the flattening of the Milky Way. This was followed by \cite{majewski_et_al_2003} who studied the great circles of M-giants and found a difference between the stream plane in the northern and southern Galactic hemispheres. \cite{johnston_et_al_2005} used this change in stream plane to measure the flattening of the Milky Way potential, via comparison both to N-body simulations and orbit integrations in flattened potentials. Finally, \cite{belokurov_sgr_precess} extended these results by measuring the pole of the stream in several locations. While this plane precession could be due to the torque from a flattened halo, as described in this work, we note that if the
Sgr dwarf galaxy had a stellar disk, the resultant stream would appear to
``precess'' even in a spherical host
\citep{penarrubia_sgr_disk,gibbons_disk}. Such an apparent debris
plane evolution is not due to the gravitational torques acting on the
stream (there are none!), but rather due to the torques experienced by
the progenitor's disk - these can exist even in a spherically
symmetric case, if the stellar disk and the progenitor's orbit are
misaligned. 

Taking advantage of a more comprehensive framework for the orbital
angular momentum evolution presented here, we show that dissipation of
the stellar stream debris can now be straightforwardly described in a
variety of aspherical hosts. As far as we know, the connection between
the flattening of the potential and the width of the stream has not
been rigorously addressed in the literature. This can be contrasted
with a rapidly growing database of Galactic stellar stream
observations, where the stream width is one of the primary parameters
reported. We envisage that by modelling the stream width it would be
possible to better characterize the shape of the dark matter potential
in the Milky Way. Such inference, of course, relies on an estimate of
the mass of the stream's progenitor - the chief factor governing the
initial spread of the debris orbital planes. On most occasions, such
an estimate is not readily available, as the majority of streams
detected so far do not have a known progenitor. Therefore, the
conventional rule of thumb has been to compare the physical width of
the stream in parsecs to the typical extent of possible progenitors,
and thus classify streams with widths under $\sim$ 100 pc as those
emanating from globular clusters and the wider ones as those from
dwarf galaxies. Here, we attempt to address the question of the
initial debris spread and show that some of the established intuition
could perhaps be misleading.

The simple picture developed here illustrates the information contained in a stream: both the precession of the stream plane and the stream's width can be used to extract the shape of the host galaxy. However, the analytic model used in this work is based on circular orbits and only captures the average evolution of these quantities. In order to constrain the shape of a galaxy using realistic streams on eccentric orbits, stream modelling is necessary \citep[e.g.][]{gibbons_et_al_2014,sanders2014,bovy2014,price_whelan_et_al_rewinder,kuepper_et_al_2015,fardal2015,bowden_gd1}.

This paper is organized as follows. In \Secref{sec:precession_rate} we
derive the precession and nutation rates of orbits in axisymmetric
potentials. In \Secref{sec:stream_width} we show that the precession
along the stream matches the precession of the orbits. Based on the
precession rate of individual orbits, we then build a model for the
evolution of the stream width and find good agreement with
simulations. In \Secref{sec:triaxial} we study orbits in triaxial
potentials and find that short and long axis loops exhibit similar
behaviour to orbits in axisymmetric potentials and therefore that the
same intuition applies. We discuss observational consequences of this
model, as well as implications for planes of satellites in
\Secref{sec:discussion}, before concluding in \Secref{sec:conclusion}.

\section{Precession and Nutation of Orbits in Axisymmetric Potentials} \label{sec:precession_rate}

Before an analytic prescription for the growth rate of stellar stream
widths in axisymmetric potentials can be devised, the behaviour of
individual orbits must be modelled. For potentials which are mildly
flattened, we will show that orbital planes effectively precess and
nutate \citep[see also][]{BT08} and derive the average precession and
nutation rate in two limits.

\subsection{Orbits near the symmetry plane in a potential with arbitrary flattening} \label{sec:equatorial_derivation}

The first limiting case we consider concerns a perturbed circular
orbit in the equatorial $x$-$y$ plane with an initial radius of
$r=R_0$ and an angular frequency $\Omega$ in a potential with
arbitrary flattening. In the following discussion we will assume that
the potential is an arbitrary function $\Phi(r)$ of flattened radius
$r$, where $r=\sqrt{x^2+y^2+z^2/q^2}$. This flattening in the potential, as opposed to the density, is a simplification. However, since the derivation is only performed for near circular orbits, it is justified. In \Appref{sec:general_expressions} we present expressions for the precession and nutation rates in a more general axisymmetric potential. The orbit is perturbed in the
$z$ direction by $\Delta z(t)$. We could also examine perturbations in
the $R$ and $\phi$ directions, but at leading order these do not
affect the orbital plane precession rate since they are not coupled to
the vertical oscillation. The equation of motion in the $z$ direction
is
\eq{ \ddot{z} = - \Phi'(r) \frac{z}{q^2 r} .}
Expanding this to first order in $z=\Delta z(t)$ we obtain
\eq{ \Delta \ddot{z}(t) &= -\Phi'(R_0) \frac{\Delta z(t)}{q^2 R_0} , \nln &= -\frac{\Omega^2}{q^2} \Delta z(t) .}
The corresponding solution is
\eq{ \Delta z(t) = \Delta z_0 \cos(\frac{\Omega}{q} t + \alpha) , \label{eq:vertical_oscillation}}
where $\Delta z_0$ is the amplitude of the perturbation and $\alpha$
is a phase, giving us a vertical frequency of $\Omega/q$. Without loss
of generality, we can set $\alpha=0$ and derive the components of the
orbital angular momentum:
\eq{ L_x &= -R_0 \Omega \Delta z_0 \Big(\frac{1}{q} \sin(\Omega t)  \sin(\frac{\Omega}{q}t ) + \cos(\Omega t) \cos(\frac{\Omega}{q}t ) \Big), \nln L_y &=  R_0\Omega \Delta z_0 \Big(\frac{1}{q}\cos(\Omega t) \sin (\frac{\Omega}{q}t)  - \sin(\Omega t) \cos(\frac{\Omega}{q}t)  \Big) ,\nln L_z &= R_0^2 \Omega .}
The leading order corrections to the radial and azimuthal positions
would give leading order corrections to $L_z$ but otherwise leave
$L_x$ and $L_y$ unaffected. Hence radial and azimuthal perturbations
do not affect the precession or nutation rates at leading
order. Re-writing the first two components of the angular momentum in
a more suggestive form we get

\eq{ L_x &= -\frac{R_0 \Omega \Delta z_0}{2} \Big( q_+ \cos(q_- \Omega  t) - q_- \cos(q_+ \Omega  t) \Big) , \nln L_y &=  \frac{R_0 \Omega \Delta z_0}{2}  \Big(   q_+ \sin(q_- \Omega  t) + q_- \sin(q_+\Omega  t)  \Big) , \label{eq:L_equatorial}}
where we have introduced $q_- = q^{-1}-1$ and $q_+ = q^{-1}+1$. If $q$
is near 1, \eqref{eq:L_equatorial} can be broken up into slowly
and rapidly oscillating terms. We see the dominant behavior comes from
the terms proportional to $q_+$ in \eqref{eq:L_equatorial} which gives
a precession frequency of
\eq{ \omega_{\rm precess} &= -\Omega q_- \nln &= \Omega (1-q^{-1}) .\label{eq:precess_near_plane} }
In addition to this precession, we also see that the angular momentum is nutating. We can compute the inclination angle, $\psi$, as
\eq{ \tan^2\psi=\frac{L_x^2 + L_y^2}{L_z^2} = \frac{\Delta z_0^2}{R_0^2} \left(1 + \frac{q_+ q_-}{2} \Big( 1 -  \cos\big( \frac{2\Omega t}{q} \big) \Big) \right) . \label{eq:L_frac_equatorial}}
This shows that the tilt of the angular momentum from the symmetry axis will nutate as it precesses, with a frequency of
\eq{ \omega_{\rm nutate} = \frac{2 \Omega}{q} . \label{eq:nutation_freq_near_plane} }
Using \eqref{eq:L_frac_equatorial}, we can also determine the amplitude of the nutation. We see that the inclination angle varies from $\frac{\Delta z_0}{R_0}$ to $\frac{1}{q} \frac{\Delta z_0}{R_0}$. Thus, for $q=0.9$ we would expect the inclination to vary by $\sim 10\%$.

\subsection{Orbits with an arbitrary orientation in a potential with small flattening}

\begin{figure}
\centering
\includegraphics[width=0.3\textwidth]{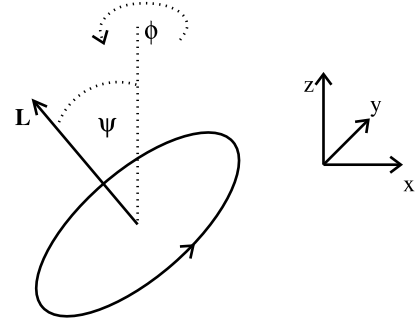}
\caption{Geometry of plane precession for a nearly circular orbit with
  an arbitrary orientation in a potential with small
  flattening. $\psi$ measures the angle between the orbit's angular
  momentum, $\mathbf{L}$, and the $z$ direction. $\phi$ measures the
  angle between the $x$ direction and the projection of the angular
  momentum in the symmetry plane spanned by $x$ and $y$.}
\label{fig:precession_geometry}
\end{figure}

The second limiting case in which we can compute the precession rate
involves a nearly circular orbit with an arbitrary inclination in a
potential with a small flattening. The geometry of this orbit is shown
in \Figref{fig:precession_geometry} along with the definition of some variables. For sufficiently small flattening
$q$, the orbit is approximately circular. In this limit, we can
integrate the torque over a single orbit and use it to compute the
precession rate of the angular momentum vector.

Once again, we consider a potential $\Phi(r)$ where $r =
\sqrt{x^2+y^2+z^2/q^2}$. We expand the radius to leading non-trivial
order in $z$ as follows: \eq{ r &= \sqrt{x^2+y^2+z^2 + z^2(q^{-2}-1)}
  \nln &= \sqrt{r_0^2 + z^2(q^{-2} -1)} \nln &= r_0 +
  \frac{z^2}{2r_0}\epsilon_q , }
where $r_0 = \sqrt{x^2+y^2+z^2}$ and we have condensed the $q$ dependence into
\eq{ \epsilon_q = q^{-2} - 1 .\label{eq:epsilon_q}}
This expansion is valid when $\frac{z^2}{r_0^2}|\epsilon_q| \ll 1$,
which requires that the potential is sufficiently spherical and that
the orbit is sufficiently close to equatorial. We note that even for a
polar orbit, i.e. $z \sim r_0$, if $q=0.95$, this expansion parameter
is 0.1 and the expansion is justified. We then use this approximate
behaviour of the flattened radius to expand the potential to leading
order as
\eq{ \Phi(r) = \Phi(r_0) + \Phi'(r_0) \frac{z^2}{2r_0} \epsilon_q . \label{eq:potential_q_expansion}}
The acceleration in this potential is given by

\eq{
\mathbf{a} = -\Big[\Phi'(r_0)\left(1-\frac{z^2}{r_0^2}\epsilon_q\right)+\Phi''(r_0)\frac{z^2}{2r_0}\epsilon_q\Big]\hat{\mathbf{r}}_0-\frac{\epsilon_q}{r_0}\Phi'(r_0)\mathbf{z}.
}

Given this acceleration, we can now compute the change in angular momentum, $\Delta \mathbf{L}$, over an orbit by integrating the torque:
\eq{\Delta \mathbf{L} = \int \mathbf{r}_0 \times \mathbf{a} \mathrm{d}t =-\frac{\epsilon_q}{r_0}\Phi'(r_0)\int(\mathbf{r}_0 \times\mathbf{z})\mathrm{d}t.}
The change in angular momentum after an orbital period $2\pi/\Omega$ is given by
\eq{ \Delta L_x &= \pi r_0^2 \epsilon_q \Omega \sin\phi \cos\psi_0 \sin\psi_0 , \nln \Delta L_y &= -\pi r_0^2 \epsilon_q \Omega \cos\phi \cos\psi_0 \sin\psi_0 , \nln \Delta L_z &= 0 ,\label{eq:delta_L}}
where we have used $\Phi'(r_0) = r_0 \Omega^2$. In order to understand how this relates to the precession we consider the orbit's initial angular momentum:
\eq{ L_x &= r_0^2 \Omega \sin\psi_0 \cos\phi , \nln L_y &= r_0^2 \Omega \sin\psi_0 \sin\phi , \nln L_z &= r_0^2 \Omega \cos\psi_0 . \label{eq:L_init}}
If the change in angular momentum over an orbital period is small compared to the angular momentum, we can equate this change to the average time derivative of the angular momentum times the orbital period, i.e.
\eq{\Delta \mathbf{L} = \left\langle \frac{d \mathbf{L}}{dt} \right\rangle \frac{2\pi}{\Omega} . \label{eq:L_equate_change}}
Plugging \eqref{eq:delta_L} and the time derivative of \eqref{eq:L_init}, allowing only for variations in $\phi$, into \eqref{eq:L_equate_change}, we see that the torque integrated over an orbit causes the angular momentum to precess at an average rate of
\eq{ \dot{\phi} = -\frac{\Omega \epsilon_q \cos\psi_0}{2} , }
resulting in a precession frequency of
\eq{ \omega_{\rm precess} = \frac{\Omega}{2} (1 - q^{-2}) \cos\psi_0 , \label{eq:precession_rate}}
where we have replaced $\epsilon_q$ with its dependence on $q$ for clarity. As mentioned above, this derivation is only valid if the orbit does not noticeably precess over a single orbit, i.e. $\omega_{\rm precess} \ll \Omega$. We note that at leading order in $(1-q)$, this gives the same result as \eqref{eq:precess_near_plane}.

We also note that this precession rate is similar to that reported in \cite{steiman_durisen_1990} which used a slightly different potential expansion that is only valid for logarithmic potentials. The $q$ dependence of their precession rate can be found in Section 5 of \cite{ibata_et_al_2001} and has the same dependence at leading order in $(1-q)$ as \eqref{eq:precession_rate} but differs at higher order.

\subsubsection{Nutation}

The nutation frequency can be qualitatively derived by considering the torque on the particle during an orbit. Looking at \Figref{fig:precession_geometry} and taking the angular momentum, $\mathbf{L}$, to be in the $x$-$z$ plane we see that a torque in the $x$ direction will cause the angular momentum to nutate while a torque in the $y$ direction will cause it to precess. While the particle is on the section of the orbit with $z>0$, the torque in $x$ undergoes a full period. The same is true for the section of the orbit with $z<0$. Thus, the nutation undergoes two periods for every vertical period. Since the vertical frequency is given by $\Omega/q$, as we saw in \eqref{eq:vertical_oscillation}, we can thus conclude that nutation frequency is given by
\eq{ \omega_{\rm nutate} = \frac{2\Omega}{q} , \label{eq:nutation_freq}}
just as in the near-equatorial case.

It is also possible to derive an approximation for the amplitude of the nutation that was described in \Secref{sec:equatorial_derivation}. We take the same setup as in the previous paragraph, i.e. the angular momentum pointed in the $x$-$z$ plane, and compute the change in angular momentum during half of the section of the orbit with $z>0$. During this quarter orbit, we find
\eq{ \Delta L_x &= \frac{1}{2} r^2 \epsilon_q \Omega \sin \psi_0 , \nln \Delta L_y &= -\frac{1}{4} \pi r_0^2 \epsilon_q \Omega \cos\psi_0 \sin\psi_0 , \nln \Delta L_z &= 0 . \label{eq:delta_L_nut}}
The change in $\Delta L_y$ causes the angular momentum to precess but the change in $\Delta L_x$ causes the angular momentum to nutate. We can then compute the change in inclination angle using the relation
\eq{ \tan \psi = \frac{\sqrt{L_x^2+L_y^2}}{L_z} .}
Expanding the right side of this equation to leading order using the changes in \eqref{eq:delta_L_nut}, and expanding $\psi$ as $\psi_0 + \delta \psi_{\rm nut}$, we find
\eq{ \delta \psi_{\rm nut} = \frac{|\epsilon_q|}{2} \cos\psi_0 \sin\psi_0 . \label{eq:nutation_amp}}
This expression is the amplitude of the nutation since after this section of the orbit, the next half with $z>0$ will have the opposite torque in the $x$ direction which will cause the orbital plane to nutate back to its initial inclination angle, albeit having precessed slightly.

In \Appref{sec:general_expressions}, we present our results for the precession and nutation rates for a more general form of the potential to make them more readily accessible for the reader.

\subsection{Comparison with orbits} \label{sec:orbit_comparison}

Now that we have built a simple model for the angular momentum
evolution, we can compare its performance with the results of direct
orbit integration in axisymmetric potentials. We use a logarithmic
potential with a circular velocity of $220$ km/s. We consider
near-circular orbits as well as eccentric orbits which start on the
$x$-axis at a position of $(30,0,0)$ kpc. For the circular orbits, the
magnitude of the initial velocity was $v_i = 220$ km/s. For each
inclination angle, $\psi_0$, the initial velocity is given by $(0,v_i
\cos \psi_0, v_i \sin \psi_0)$. For the eccentric orbits, the
magnitude of the initial velocity is given by the tangential velocity
required to have a pericentre at $15$ kpc given an apocentre of $30$
kpc in a spherical potential. This sets the magnitude of the velocity
to be $v_i = 149.55$ km/s.

\begin{figure}
\centering
\includegraphics[width=0.5\textwidth]{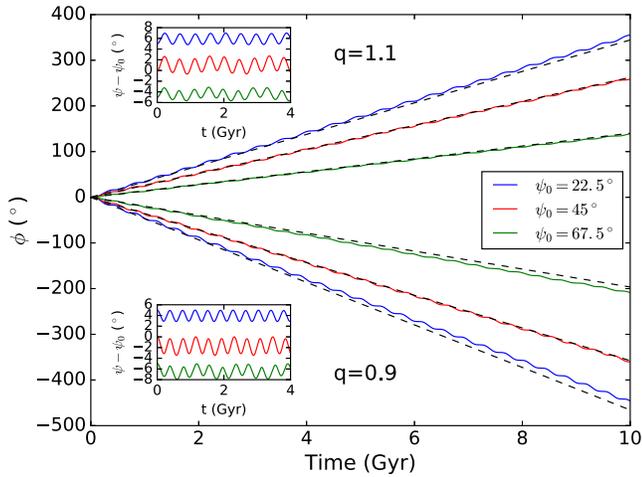}
\caption{Evolution of the precession and nutation angles for near
  circular orbits with different inclinations. The main plot shows the
  precession angle for three different inclination angles in
  potentials with $q=1.1$ and $q=0.9$. The colored curves show the precession and nutation measured from orbits and the dashed-black curves show the expected precession from \protect\eqref{eq:precession_rate}. As the orbit is inclined
  farther away from the symmetry axis, the precession rate decreases,
  as expected from \protect\eqref{eq:precession_rate}. Note also that
  the sense of precession reflects the sense of flattening of the
  potential, i.e. whether it is prolate, with $q>1$, or oblate, with
  $q<1$. For orbits in prolate potentials, the orbital plane precesses
  in the same direction as the orbital motion. The opposite is true
  for orbits in oblate potentials. The insets show the evolution of
  the nutation angle. We see that the nutation appears almost
  sinusoidal for these near circular orbits. It is also evident that
  the nutation frequency is not sensitive to the inclination angle, as
  expected from \protect\eqref{eq:nutation_freq}. }
\label{fig:phi_angle_var_circular}
\end{figure}

\Figref{fig:phi_angle_var_circular} shows the precession and nutation of
near-circular orbits with three different values of $\psi_0$ and two
different values of $q$. It is demonstrated that the overall dominant
motion is that of precession since the nutation only rocks the orbital
plane back and forth and does not lead to any secular
trends. Additionally, it is clear that orbits precess at a faster rate
as their planes are tilted closer to the symmetry axis, i.e. as we
decrease $\psi_0$. Note that the sense of the precession changes as
the potential is switched from the oblate one, i.e. with $q<1$, to a
prolate one, i.e. with $q>1$. As far as the nutation amplitude is
concerned, it is largest for the orbits with $\psi_0 = 45^\circ$ and
reaches similar values for those with $\psi_0=22.5^\circ$ and
$\psi_0=67.5^\circ$, as expected from
\eqref{eq:nutation_amp}. Importantly, the model precession rates given
by \eqref{eq:precession_rate} match the actual precession rates quite
well.

\begin{figure}
\centering
\includegraphics[width=0.5\textwidth]{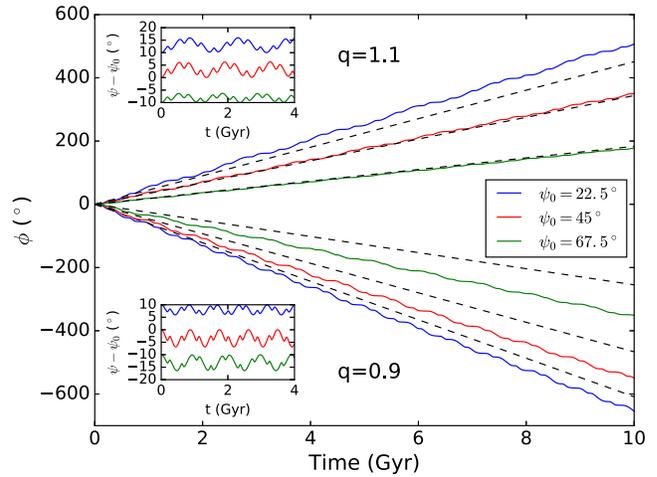}
\caption{Evolution of the precession and nutation angles for eccentric
  orbits (with pericentre at 15 kpc and apocentre at 30 kpc) with different
  orbital inclinations. This Figure is similar to
  \protect\figref{fig:phi_angle_var_circular} so we will only
  highlight the differences. For example, note that the match of the
  precession rate is not as good as for circular orbits, especially
  for $q=0.9$. In addition, the nutation exhibits a more complex
  behavior with oscillations on varying time scales.}
\label{fig:phi_angle_var}
\end{figure}

\Figref{fig:phi_angle_var} presents the behaviour of precession and
nutation angles for an eccentric orbit as a function of time. Note
that our prediction for the precession rate is not quite as good as in
the circular orbit case. However, the overall trends still hold with
the precession rate decreasing as we tilt the orbital plane away from
the symmetry axis. Compared to the circular orbit case, the nutation
appears to be more complicated with a larger amplitude and pronounced
fluctuations on several different time scales. As before, the dominant
evolution of the angular momentum is precession, with the nutation
only causing an oscillation of the orbital plane.

\begin{figure}
\centering
\includegraphics[width=0.5\textwidth]{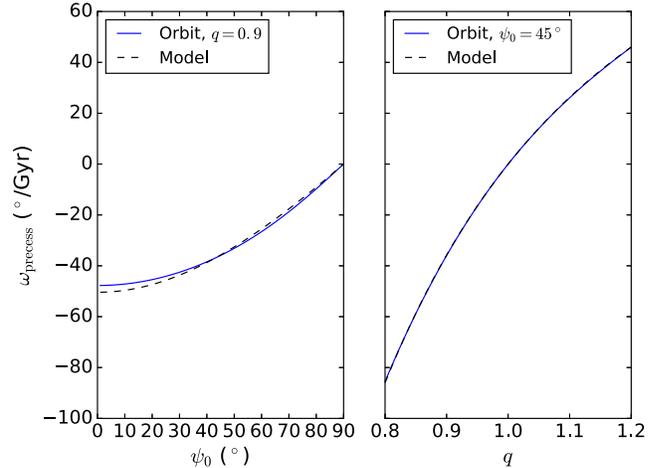}
\caption{Precession rates for circular orbits as a function of
  inclination angle and flattening. Reassuringly, our model (dashed
  black line) matches the results of orbit integration (solid blue
  line) quite well. The model curve is given by
  \protect\eqref{eq:precession_rate} where $\Omega$ is measured from
  the orbit. }
\label{fig:precession_rate}
\end{figure}

\Figref{fig:precession_rate} reports the comparison between the actual
precession rate and our model as a function of the orbital inclination
and the potential flattening. This test is carried out for the
near-circular orbits described above. Clearly, the model captures the
precession rate dependence on the inclination angle rather well. In
addition, the model matches the dependence on the flattening
remarkably well. We note that the comparison for varying flattening
values was made for a particular inclination angle, namely
$\psi_0=45^\circ$. As we can see from the left panel of
\Figref{fig:precession_rate}, this is very close to the angle where the
match is best. If we had chosen another angle, the match would
evidently not have looked as good.

\begin{figure}
\centering
\includegraphics[width=0.5\textwidth]{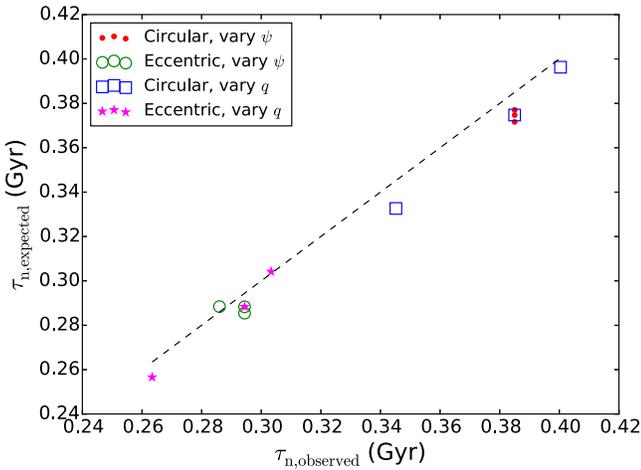}
\caption{Comparison of the nutation period found in orbit integration
  and the expected nutation period stipulated by
  \protect\eqref{eq:nutation_freq}. The observed nutation is computed
  period by taking Fourier transforms of the nutation pattern of
  near-circular and eccentric orbits and identifying the dominant
  frequency. Note that for the eccentric orbits, we have neglected the
  slow oscillation which is not captured by our model. For the setups
  where we varied $\psi$ we took $q=0.9$ and $\psi_0 \in [22.5^\circ,
    45^\circ, 67.5^\circ]$. For the setups where we varied $q$ we took
  $\psi_0=45^\circ$ and $q \in [0.8,0.9,0.95]$.}
\label{fig:nutation}
\end{figure}

\Figref{fig:nutation} quantifies the performance of the model
describing the nutation frequency. In order to gauge the orbital plane
nutation, we take Fourier transforms of the nutation angle, as shown
in the insets of \Figref{fig:phi_angle_var_circular} and
\Figref{fig:phi_angle_var}, and identify the dominant frequency. This
is then converted into a period and compares against the nutation
period predicted by our model, i.e. \eqref{eq:nutation_freq}. We considered five setups where we first
kept $q=0.9$ fixed and varied $\psi_0 \in [22.5^\circ, 45^\circ,
  67.5^\circ]$. Then we kept $\psi_0 = 45^\circ$ fixed and varied
$q\in[0.8,0.9,0.95]$. We considered both near-circular and eccentric
orbits. We see that the nutation period varies with $q$ but is not
very sensitive to $\psi_0$, as expected from
\eqref{eq:nutation_freq}. We note that for the eccentric orbits we
have neglected the slowly varying oscillation and instead show the
period corresponding to the high-frequency nutation. The slower
nutation for the eccentric orbits is not captured by our simple model
so we cannot make any quantitative comparison.

\begin{figure}
\centering
\includegraphics[width=0.5\textwidth]{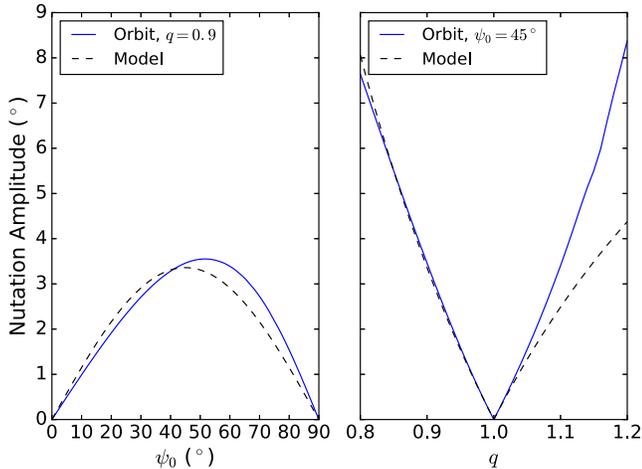}
\caption{Nutation amplitude for near-circular orbits as a function of
  inclination angle and flattening. The model, given by
  \protect\eqref{eq:nutation_amp} (dashed black line) matches the
  overall behaviour gleaned from orbit integration (solid blue
  line). For example, the model reproduces the overall trend in
  inclination angle: the nutation amplitude is zero for polar and
  equatorial orbits and increases as we move away from these
  cases. The model also captures the dependence on flattening for
  orbits in oblate potentials. However, it does not do as well for
  orbits in prolate potentials.}
\label{fig:nutation_amp}
\end{figure}

Finally, \Figref{fig:nutation_amp} examines the behaviour of the
nutation amplitude for near-circular orbits as a function of the
orbital plane inclination and potential flattening. The nutation
amplitude found in simulations is compared with that given by our
model. As evidenced by the Figure, the model reproduces the overall
trend in inclination angle: polar and equatorial orbits have zero
nutation amplitude, and the nutation amplitude increases as the orbit
tilts away from these boundary cases. Note that while the model
reproduces the nutation amplitude of orbits in oblate potentials, it
performs less well for prolate potentials. It is unclear why our model
fails in this particular way, especially given how well it reproduces
the dependence of the precession rate on the flattening as shown in
\Figref{fig:precession_rate}.

\section{Evolution of streams in axisymmetric potentials} \label{sec:stream_width}

\Secref{sec:precession_rate} develops a simple model which can capture
the precession and nutation rates together with the nutation amplitude
for orbits in axisymmetric potentials. With this foundation in place,
it is now straightforward to describe the evolution of streams in
these potentials. First, we will investigate how tidal debris planes
precess. Since streams are made up of stars on a variety of orbits,
the streams themselves do not necessarily need to precess in the same
way an orbit would. However, here we confirm that i) for the
progenitors we considered ii) in a logarithmic potential, the stream
debris plane does indeed precess like the progenitor's orbital
plane. Next, we build a simple model which captures how stream widths
grow in axisymmetric potentials. This model does not rely on how the
stream itself precesses, but instead relies on how the orbital planes
of individual stars in a stream precess.

\subsection{Streams precess like their progenitor's orbit}

\begin{figure}
\centering
\includegraphics[width=0.5\textwidth]{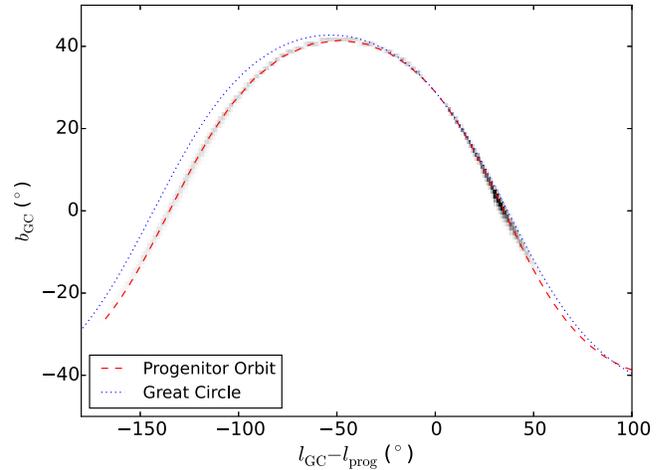}
\caption{Progenitor's orbit (dashed red line) and the stream as viewed
  from the galactic centre. The dotted blue line shows the
  Galactocentric great circle defined by the current angular momentum of the
  progenitor. The 2D greyscale histogram shows the density of stream stars
  on the sky where particles within 2 kpc of the progenitor have been
  masked out. This stream was evolved on an eccentric orbit on an
  eccentric orbit in a logarithmic potential with a flattening of
  $q=0.9$ and $\psi_0 = 45^\circ$. The progenitor is at $l_{\rm
    GC}-l_{\rm prog} = 0$. }
\label{fig:orbit_stream_sky}
\end{figure}

To generate a template tidal stream, we evolve a globular cluster
modelled by a King profile on the eccentric orbit described in
\Secref{sec:orbit_comparison} with $q=0.9$ and $\psi_0 =
45^\circ$. The details of the simulations are given in
\Secref{sec:nbody_sims}. \Figref{fig:orbit_stream_sky} gives the view
of the stream's stellar density on the sky as observed from the centre
of the Galaxy. According to the Figure, the stream does not follow the
great circle (shown in blue) defined by the normal to the current
angular momentum of its progenitor. This is a clear illustration of
the evolution of the stream debris plane due to precession.

\begin{figure}
\centering
\includegraphics[width=0.5\textwidth]{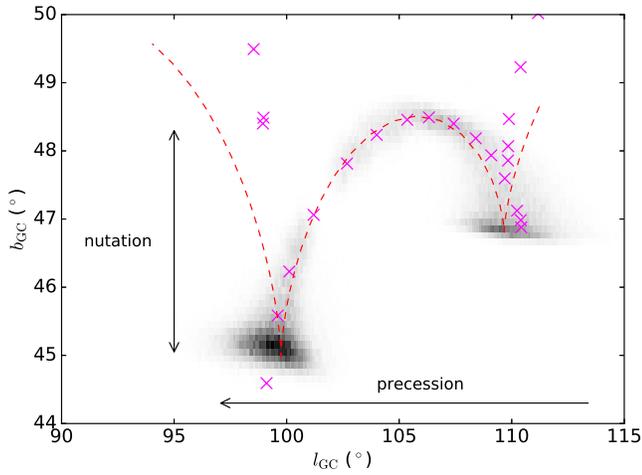}
\caption{Evolution of the pole of the progenitor's orbit as compared
  to that of the pole of the stream plane in Galactocentric
  coordinates. The dashed red line shows the pole of the progenitor's
  orbit. The grey histogram shows the density of stream poles as
  defined by the angular momentum of each particle and particles
  within 2 kpc of the progenitor have been masked out. The magenta
  crosses show the best-fit planes to segments of the stream which are
  10$^\circ$ long as seen from the galactic centre. The orbit and
  stream precess in the same way. Thus the same intuition can be
  applied to explain the evolution of both. This stream was evolved on
  an eccentric orbit in a logarithmic potential with a flattening of
  $q=0.9$ and $\psi_0 = 45^\circ$. Note that the length of the lines with arrows denoting
  the precession direction and nutation amplitude show predictions from our model for the amount of
  precession and the nutation amplitude expected for the simulated stream length. }
\label{fig:orbit_stream_plane}
\end{figure}

\begin{figure}
\centering
\includegraphics[width=0.5\textwidth]{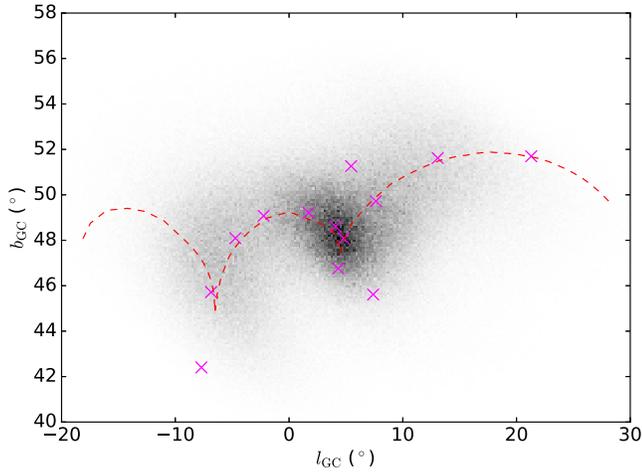}
\caption{Similar to Figure~\ref{fig:orbit_stream_plane} but for a more
  massive progenitor with $M=10^8 M_\odot$ and $r_c = 500$ pc. As with
  the less massive progenitor shown in
  \protect\figref{fig:orbit_stream_plane}, the orbit and stream
  precess in the same way. This stream was evolved in a logarithmic
  potential with a flattening of $q=0.9$ and $\psi_0 = 45^\circ$. }
\label{fig:orbit_stream_plane_1e8Msun}
\end{figure}

\Figref{fig:orbit_stream_plane} examines the motion of the debris pole
by comparing the track of the angular momentum of the progenitor along
its orbit and the track of the normal of the debris material in the
stream at a single epoch. Evidently, the angular momentum of the
stream follows the angular momentum of the progenitor's orbit. Thus,
it is justified to extend the explanation of the precession of the
orbit to the precession of the stream. Note that according to the
Figure, the stream debris pole inferred from fitting planes to
portions of the debris along the stream (shown in magenta crosses)
agrees well with the orbital angular momentum. Thus, a Galactocentric
observer would be able to infer the stream angular momentum evolution
without proper motion or distance measurements. Of course, for an
observer at the Sun, distances to the stream will be required to
disentangle the effects of the Galactic parallax and the gravitational
torques. \Figref{fig:orbit_stream_plane_1e8Msun} is an analogue of
\Figref{fig:orbit_stream_plane} for a more massive progenitor with
$M=10^8 M_\odot$ and $r_c = 500$ pc. Once again, it is clear that the
stream precesses and nutates just like the orbit of its progenitor,
albeit with a more pronounced scatter.

\subsection{Model for the stream width growth} \label{sec:stream_width_growth}

To build a model of the tidal debris dissipation, a stream is
considered to be made up of stars which have been stripped from a
common progenitor, and are now following nearby orbits in the host
galaxy potential. The width of the stream is governed by the rate at
which these stars move away from the progenitor's orbital plane. In a
spherical potential, stars in the stream would oscillate about the
progenitor's plane, creating a stream with a near-constant
width. However, in an axisymmetric potential, the orbital plane of
each star will precess at a rate governed by
\eqref{eq:precession_rate}. Since each star is on a slightly different
orbit and has a slightly different tilt relative to the symmetry axis,
the orbital plane of each star will precess at a different rate
causing the stream to fan out.

\begin{figure}
\centering
\includegraphics[width=0.45\textwidth]{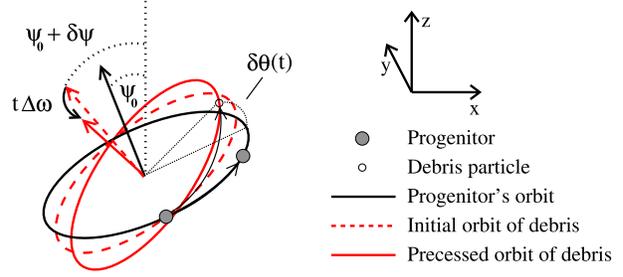}
\caption{Orbital planes of the progenitor and a stripped particle,
  each inclined by a different angle relative to the symmetry
  plane. The stream progenitor's initial position is represented by a
  filled grey circle near the bottom of the figure and its initial
  orbit is given by the solid black line. The initial orbit of a
  particle which has just been stripped is shown by the dashed red
  line. The progenitor's orbit is tilted by $\psi_0$ relative to the
  symmetry axis and the particle's orbit is tilted by $\psi_0 + \delta
  \psi$. The progenitor and stream particle precess at rates governed
  by their orbital inclination,
  i.e. \protect\eqref{eq:precession_rate} - this differential plane
  precession is what causes the stream to fan out. The solid red line
  shows the orbital plane of the particle at time $t$ since stripping,
  after which it has precessed by $t \Delta \omega$
  \protect\eqref{eq:differential_precession_rate} around the $z$-axis
  relative to the plane of the progenitor. The short thin black solid
  line shows the trajectory of the particle in the frame where we keep
  the progenitor's plane fixed, and the particle's location at time
  $t$ is given by the small empty circle. The progenitor is also shown
  at this later time with the filled grey circle on the right. $\delta
  \theta(t)$ is the angle between the particle's location and the
  orbital plane of the progenitor for an observer in the centre of the
  potential.}
\label{fig:stripping}
\end{figure}

The stream growth rate can be computed as follows. Let us start with a
progenitor whose orbit is tilted by $\psi_0$ with respect to the
symmetry axis. We clarify the geometry of this setup in
\Figref{fig:stripping} where the orbital plane of the progenitor and
the orbital plane of a stream particle which has just been stripped
are both sown. The difference in their tilt, $\delta \psi$, is
governed by the ratio of the velocity component of the stream particle
which is perpendicular to the progenitor's orbital plane to the
velocity along the orbit. Naturally, the velocity along the orbit is
dominated by the progenitor's velocity. Since stream particles are
released at the Lagrange point, we take the width of the velocity
distribution to be the velocity dispersion of the progenitor at the
tidal radius. Furthermore, we assume that the stars are stripped at
pericenter, when the tidal radius is the smallest. This tells us that
the spread in orbital plane angles at stripping is approximately
\eq{ \Delta \psi \sim \frac{\sqrt{\frac{Gm}{3 r_{\rm tidal}}}}{v_{\rm peri}} , \label{eq:psi_sigma}}
where $v_{\rm peri}$ is the magnitude of the progenitor's velocity at pericenter.

The setup presented in \Figref{fig:stripping} places the progenitor on
the $-y$ axis to simplify the subsequent derivation. The particle and
the progenitor are both taken to be on a circular orbit within their
planes with an orbital frequency $\Omega$. The precession rates of
these planes are given in \eqref{eq:precession_rate} and the
difference in their precession rate is given by
\eq{ \Delta \omega_{\rm precess} = \frac{\Omega \epsilon_q}{2} \sin\psi_0 \delta \psi , \label{eq:differential_precession_rate}}
where we have assumed that $\delta \psi \ll \psi_0$, i.e. that the
velocity dispersion at the tidal radius is much smaller than the
velocity at pericentre.

The angle between the stream particle and the progenitor's orbit on
the sky for an observer located at the centre of the potential, can be
arrived at through a series of rotations. We start with the two orbits
in the $x$-$y$ plane, both parametrized as $(r_0 \sin \Omega t, - r_0
\cos\Omega t, 0)$. We then rotate each orbit in the $+y$ direction
(i.e. into the page in \figref{fig:stripping}) by their
respective (initial) inclination angles, i.e. $-\psi_0$ and $-(\psi_0
+ \delta \psi)$. Next, we rotate by the star's differential precession
angle, $\Delta \omega_{\rm precess} t$, in the $+z$
direction. Finally, we rotate by $\psi_0$ in the $-y$ direction.
After these rotations, the progenitor's orbit returns back to the
$x$-$y$ plane while the star acquires an angular offset $\delta
\theta(t)$. Adding the effect of the three rotations gives the polar
angle between the progenitor's orbit and the stream particle is given
by
\eq{ \delta \theta(t) = \left(-\sin \Omega t + \frac{\epsilon_q \Omega t}{2} \sin^2 \psi_0 \cos \Omega t \right)\delta \psi\label{eq:delta_theta_sky} .}
Note that even if the potential is spherical, i.e. $\epsilon_q = 0$,
there is still a variation of the width over time. The reason for this
is clear from \Figref{fig:stripping}: even if angular momentum is
conserved, orbits on these two planes will periodically meet at the
nodes between the planes, and the angle between them will vary
sinusoidally. The $\psi_0$ dependence of the angle is also
unambiguous. One factor of $\sin \psi_0$ comes from the differential
precession in \eqref{eq:differential_precession_rate}. The second
$\sin \psi_0$ factor is due to the geometrical connection between the
differential precession and the relative angle between the two
orbits. For example, if the progenitor's orbit close to equatorial,
precession does not change the angle on the sky, while if the orbit is
close to polar, the change in the angle on the sky is equal to the
precession angle.

Having derived the angle on the sky of a single star relative to the
progenitor's track, it is now feasible to compute a characteristic
stream width. We compute the average of $\delta \theta^2$ and ignore
the time dependence outside of the oscillatory terms, which finally
gives a characteristic width of
\eq{ w \approx \frac{\Delta \psi}{\sqrt{2}} \sqrt{ 1 + \frac{\Omega^2 t^2 (1-q^{-2})^2 \sin^4 \psi_0}{4} } . \label{eq:char_width} }
Unsurprisingly, the stream width is expected to be constant in
spherical potentials. In addition, the stream width grows faster as we
tilt away from equatorial orbits. Thus, streams generated by
progenitors on nearly polar orbits should fan out the quickest. In
\Appref{sec:general_expressions}, we present an expression for the
stream width for a more general form of the potential.

This expression above can also be used to gain insight into the
variation of the width of a stream with the progenitor's
Galactocentric radius. To simplify the discussion, we assume that the
potential is spherical and that the progenitor is on a circular
orbit. As a result, the width of the resulting stream is just
proportional to the initial angular spread, i.e. $\Delta \psi$ from
\eqref{eq:psi_sigma}. This spread is largely independent of radius
since both the velocity dispersion at the tidal radius and the
circular velocity have the same explicit dependence on the
Galactocentric distance, i.e. $\propto r^{-1/2}$. However, they have a slightly
different dependence on the enclosed mass, $M(r)$, which makes $\Delta
\psi$, and hence the stream width, a weak function of radius:
\eq{ w \propto M(r)^{-1/3}. \label{eq:width_r_dep}}
Thus, if identical progenitors are disrupted at different
Galactocentric radii, the angular widths of the resultant streams are
expected to be roughly constant, with streams farther out slightly
narrower due to the larger enclosed host's mass. As a result, streams
in the outer parts of the Galaxy will have larger physical widths,
therefore caution must be taken when using the stream's cross-section
in parsecs to infer the nature of its progenitor.

We note that this model for the stream width only holds for observers in the center of the potential. If the observer is not in the plane of the stream, the width they measure will also be affected by the spread of the debris within the stream plane \citep[e.g.][]{johnston_et_al_2001,amorisco_2014,hendel_johnston_2016}. The relative contribution of the width within and perpendicular to the plane depends on the distance and orientation of the observer relative to the stream plane. In the numerical examples presented in \Secref{sec:nbody_sims} we will discuss how these two widths compare.

\subsection{Comparison with N-body simulations} \label{sec:nbody_sims}

In this Section, the prediction for the characteristic width given by
\eqref{eq:char_width} is tested against the widths of streams
generated in N-body simulations. The simulations used for this
comparison were carried out with the N-body part of \textsc{Gadget-3},
which is similar to \textsc{Gadget-2} \citep{springel_2005}. Each
progenitor was initialized to follow a King density profile with
$M=10^5 M_\odot$, $w=5$, and $r_c = 13$ pc, i.e. similar to a typical
(if slightly puffy) globular cluster. Each cluster prototype was
represented by $10^6$ particles and a softening of 1 pc was used. The
clusters were all placed on the $x$-axis at a position of $(30,0,0)$
kpc. The clusters were then evolved on the eccentric orbits described
in \Secref{sec:orbit_comparison} for 10 Gyr in a logarithmic potential
with a circular velocity of $220$ km/s.  We ran a total of six
simulations: for a flattening of $q=0.9$, we ran a progenitor on
orbits with $\psi_0 = 22.5^\circ, 45^\circ, 67.5^\circ,$ and
$90^\circ$, in addition, for an inclination of $\psi_0 = 45^\circ$, we
ran a simulation of a disruption of a progenitor in a potential with
$q=0.8$ and $0.95$. We also considered a more massive cluster with
$M=10^8 M_\odot$ and $r_c = 500$ pc which was represented by $10^6$
particles with a softening of 38 pc. This cluster was placed on the
eccentric orbit in the same potential with $q=0.9$ and
$\psi_0=45^\circ$. This more massive cluster was used in
\Figref{fig:orbit_stream_plane_1e8Msun} to show that even streams
generated by massive progenitors precess like the progenitor's orbit.

\begin{figure}
\centering
\includegraphics[width=0.5\textwidth]{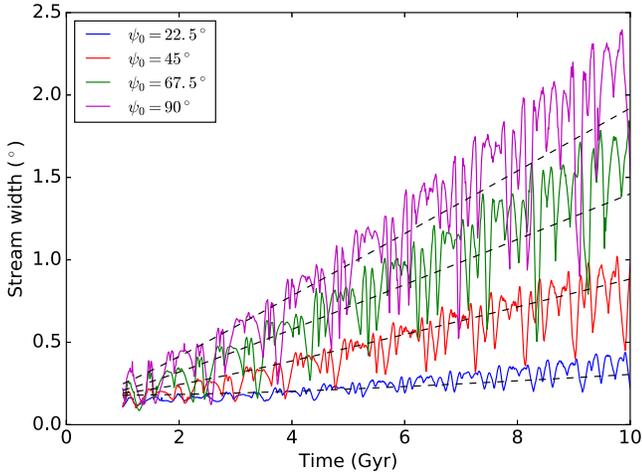}
\caption{Evolution of the maximum stream width for streams on
  eccentric orbits with varying inclination. These streams are evolved
  in a potential with $q=0.9$. The dashed black line shows our model
  given by \protect\eqref{eq:char_width}. The coloured, solid lines
  show the maximum stream width as measured from the simulations. We
  see that the stream width grows more rapidly as we increase the
  inclination angle. Note, that the oscillation of the stream width
  also grows with increasing inclination.}
\label{fig:stream_width_angle_var}
\end{figure}

\begin{figure}
\centering
\includegraphics[width=0.5\textwidth]{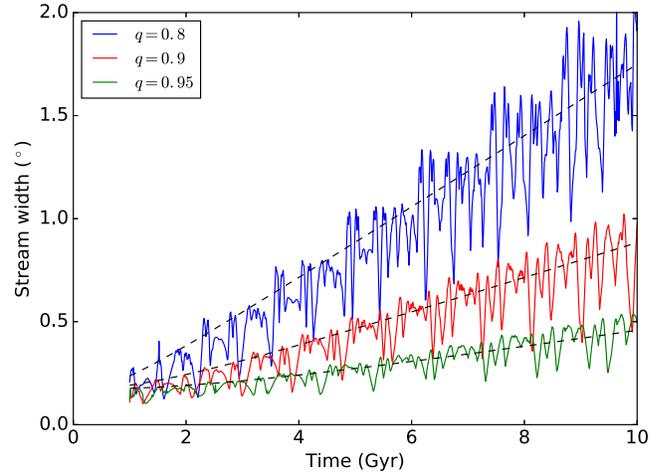}
\caption{Evolution of the maximum stream width for streams on
  eccentric orbits in potentials with varying flattening. These
  streams are evolved with an inclination angle of $\psi_0 =
  45^\circ$. The dashed black line shows our model as stipulated by
  \protect\eqref{eq:char_width}. The coloured, solid lines show the
  maximum stream width as measured from our simulations. We see that
  the stream width grows more rapidly as we make the potential
  flatter.}
\label{fig:stream_width_q_var}
\end{figure}

\Figref{fig:stream_width_angle_var} presents the time dependence of
the maximum stream width for different orbital inclinations. The width is computed by separately finding the best-fit plane to the leading and trailing arm, excluding the particles within 2 kpc of the progenitor. For each arm, the dispersion of the width is computed in 5$^\circ$ bins along the stream. The maximum width is then taken across all segments. We chose
to measure the maximum stream width since our model follows the width
of a single stripping event. As expected, the stream width grows faster as the
orbital plane is tilted farther away from the symmetry axis.  The
model prediction for the stream width is also plotted. The theoretical
curves are based on \eqref{eq:char_width} and use \eqref{eq:psi_sigma}
to estimate the dispersion in the initial orbital plane
orientation. Reassuringly, there appears to be a good match between
the model and the results of the simulations across a range of
inclinations, albeit with increasing scatter around the average width
for orbits farther away from the symmetry axis. Note that if we instead measured the average width along the stream, we would find a narrower width since this would include more recently stripped material which has had less time to fan out. To compare our model against this average width we would need to integrate over the disruption history
of the progenitor, accounting for the fanning from each stripping event.
\Figref{fig:stream_width_q_var} displays the stream width evolution
for different values of the potential flattening. Once again, the
analytic model matches the stream behaviour in the simulation,
confirming the prediction that the flatter potentials cause faster
stream scattering. Finally, we note that the massive cluster with $M=10^8 M_\odot$ has a larger maximum stream width than expected from \eqref{eq:char_width}. This is most likely because when the progenitor was initialized, there were particles outside the tidal radius and these were stripped immediately, while the model assumes that material is only stripped from the tidal radius.

As discussed in \Secref{sec:stream_width_growth}, the stream is also broadening within the stream plane due to differential apsidal precession. The width that an observer sees on the sky will depend on their orientation and distance with respect to the stream plane. In order to gauge the relative importance of the extent within and perpendicular to the stream plane for the numerical examples presented here, we can compare their widths. We repeat the maximum width calculation above, except instead of taking the maximum of the angular width, we take the maximum the physical width within the stream plane and perpendicular to the stream plane. As above, we compute the dispersion for $5^\circ$ segments as viewed from the stream plane. For the case of $q=0.9$, we find that the width within the plane is larger than the perpendicular width for $\psi=22.5^\circ$. For $\psi=45^\circ$ the widths are comparable, although the width perpendicular to the plane is slightly larger. For $\psi=67.5^\circ$ and $\psi=90^\circ$ we find that the width perpendicular to the plane is larger. For the cases when we vary $q$ we find that the width within the plane is larger for $q=0.95$ and that the width perpendicular to the plane is slightly larger for $q=0.9$ and dominates the in-plane width for $q=0.8$. Thus, we must be cautious when interpreting an observed stream width since the contributions from the debris spread within and perpendicular to the plane can be comparable, depending on the potential and the tilt of the stream plane.

\section{Extension to Triaxial Potentials} \label{sec:triaxial}

The discussion above has been restricted to axisymmetric potentials
since the orbital plane evolution and stream width can be approximated
analytically for these. However, as we show below the same intuition
applies to short and long-axis loop orbits in triaxial potentials.

Orbits in triaxial potentials are more complex than those in
axisymmetric potentials since the potential has no rotational
symmetries. As a result, none of the angular momentum components are
conserved. Insight can be garnered through inspection of orbits in the
St\"ackel potentials \citep{dZ85} which are always regular, but
generically triaxial potentials produce regions of regular orbits
separated by regions of chaotic orbits \citep{Valluri98}. Orbits in
triaxial potentials can be classified into three categories:
short-axis loops, long-axis loops, and box orbits. The short and
long-axis loops have angular momenta which are close to the short and
long axis of the potential. Although stars on these orbits do not
conserve any component of their angular momenta, they maintain the
sign of their angular momentum along their respective axes (i.e. short
or long axis). Thus, their orbital planes do not stray too far from
the symmetry axis and they look roughly like orbits in axisymmetric
potentials. Box orbits do not have this property: their motion cannot
be thought of as precessing in any sense so we cannot extend the
results of this work to this class of orbits. However, the action-based approach of \cite{pontzen_et_al_2015} may allow one to build a model for box orbits, as well as for more general aspherical potentials.

Let us explore how much of the intuition developed for orbits in
axisymmetric potentials extends to short and long axis loops. We
consider a triaxial logarithmic potential with $v_{\rm circ} = 220$
km/s and a radius of $r = \sqrt{x^2+y^2/q_y^2+z^2/q_z^2}$. We select
$q_y=0.95$ and $q_z=0.9$. Thus, $z$ is the short axis and $x$ is the
long axis. We launch ``star'' particles from the $y$-axis at (0,30,0)
kpc, with a velocity of $(-v_0 \cos\psi_0,0,v_0 \sin\psi_0)$, where
$v_0=149.55$ km/s, i.e. the same speed as our eccentric orbits
discussed previously. The angle $\psi_0$ gives the initial orientation
of the orbital plane relative to the $z$-axis in the $x$-$z$
plane. Thus, $\psi_0=0^\circ$ corresponds to a short axis loop which
circulates about the $z$-axis. Likewise, $\psi_0=90^\circ$ corresponds
to a long axis loop which circulates about the $x$-axis.

\begin{figure}
\centering
\includegraphics[width=0.5\textwidth]{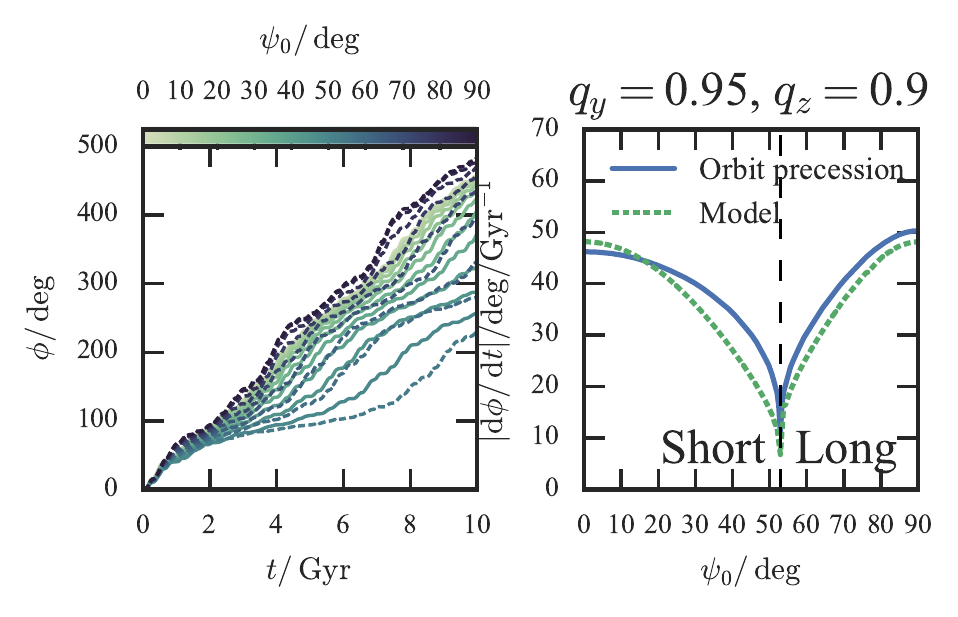}
\caption{Precession of orbits in a triaxial potential with a
  flattening of $q=0.9$ for the $z$ axis and $q=0.95$ for the $y$
  axis. $\psi_0$ is the angle between the initial angular momentum and
  the $z$-axis. Orbits with $\psi_0 < 52.9^\circ$ are short axis loops
  and orbits with $\psi_0 > 52.9^\circ$ are long axis loops. The
  precession angle, $\phi$, is always measured about the axis which
  the orbit is circulating. The left panel shows the precession angle
  as a function of time for orbits with various $\psi_0$, while the
  right panel shows the average precession rate. The solid blue curve
  on the right panel shows the precession rate determined from linear
  fits to the precession angle evolution. The dashed green curve is
  our model from \protect\eqref{eq:precession_rate} as described in
  the text. As in the axisymmetric case, we see that the precession
  rate decreases as we move away from the symmetry axis.}
\label{fig:triax_q_09509}
\end{figure}

\Figref{fig:triax_q_09509} displays the precession rate as a function
of $\psi_0$ for the orbital setup described above. The left panel of
Figure shows the precession angle $\phi$ similar to
\Figref{fig:phi_angle_var_circular} and
\Figref{fig:phi_angle_var}. The right panel of
\Figref{fig:triax_q_09509} gives the average precession rate
determined by a linear fit to the curves on the left.  As in the
axisymmetric case, the precession is fastest for orbits with angular
momentum near the symmetry axes. For orbits farther away from the
symmetry axis, the precession rate decreases. The solid blue curve in
the right panel of \Figref{fig:triax_q_09509} shows the precession
rate as determined by the orbit integration. The dashed green curve
is the expected precession rate from \eqref{eq:precession_rate}. Here
$\psi_0$ is the initial inclination angle relative to the $z$-axis, $\Omega$ is determined from the frequency of circulating about
the short or long axis, and $q$ is obtained from the ratio
$\Omega/\Omega_z$ where $\Omega_z$ is the vertical frequency along the short or long axis. This
choice is motivated by \eqref{eq:vertical_oscillation}. We find that
our model reproduces the overall trends exhibited by the orbits with
the precession rate decreasing as we move away from the long and short
symmetry axes. This is similar to the behaviour discussed in
\Secref{sec:orbit_comparison} for axisymmetric potentials.

This change in the precession rate also allows us to understand how
the stream width would grow for these orbits. As we argued in the
previous Section, the growth rate of the stream width is controlled by
the differential precession rate and a geometrical factor. Our model
predicts that the stream width should grow the fastest for the orbits
where the precession rate is the steepest function of
angle. Therefore, we expect that the stream width will grow the
slowest for orbits near the short and long symmetry axes and increase
as we move away from these axes.

\begin{figure}
\centering
\includegraphics[width=0.5\textwidth]{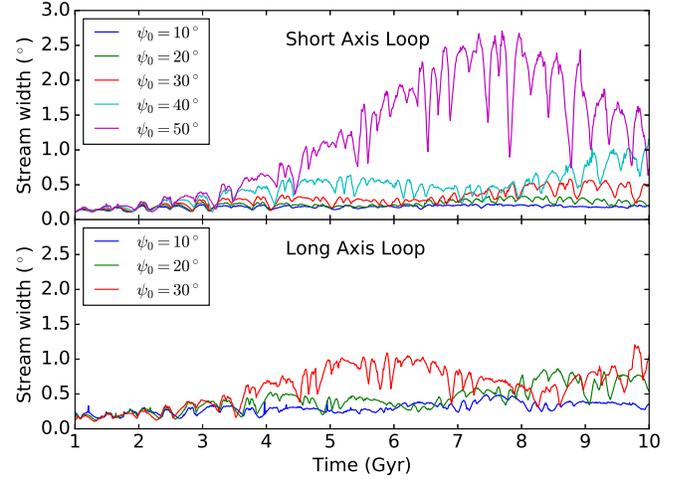}
\caption{Maximum width of streams in a triaxial potential. The setup
  is described in the text of \protect\secref{sec:triaxial}. The
  progenitors are launched from the $y$-axis in a potential with
  $q_y=0.95$ and $q_z=0.9$. The angle $\psi_0$ refers to the angle
  between the initial angular momentum vector and the $z$ and $x$-axis
  for the short and long axis loops respectively. We see that the
  further we move away from a short or long axis loop, the wider the
  stream becomes. }
\label{fig:width_triax}
\end{figure}

We can now test this prediction using N-body simulations. We take the
same King profile described in \Secref{sec:nbody_sims}, except with
$10^5$ particles instead of $10^6$, and launch them from the $y$-axis
at $(0,30,0)$ kpc with a velocity of $(-v_0 \cos\psi_0,0,v_0
\sin\psi_0)$ where $v_0=149.55$ km/s. We ran eight simulations with
$\psi_0$ ranging from 10$^\circ$-80$^\circ$ in steps of 10$^\circ$. In
\Figref{fig:width_triax}, we show how the maximum width of the streams
evolves in time. From \Figref{fig:triax_q_09509} we see that orbits
with $\psi_0 < 52.9^\circ$ are short axis loops which conserve the
sign of their angular momentum about the $z$-axis, while those with
$\psi_0 > 52.9^\circ$ are long axis loops which conserve the sign of
their angular momentum about the $x$-axis. For both classes of orbits
we list the angle relative to the axis about which they are
circulating in \Figref{fig:width_triax}. We see that the intuition
developed in the axisymmetric case carries over to this triaxial case:
the further we tilt the orbital plane away from the short and long
axis of the potential, the wider the stream becomes.

\section{Discussion} \label{sec:discussion}

\subsection{Observable consequences}

The stream plane precession and nutation, as well as the evolution of
the stream width, are all observable consequences of the results in
this work. We will now explore how these can be utilized to constrain
the flattening and orientation of the Milky Way potential.

\subsubsection{Stream centroid evolution due to precession and nutation} \label{sec:obs_precess_nut}

As Figures \ref{fig:orbit_stream_plane} and
\ref{fig:orbit_stream_plane_1e8Msun} reveal, the stream debris plane
evolves in a coherent fashion: it slowly strays around the axis of
symmetry in the host potential and swings up and down in the
perpendicular direction. Moreover, as evidenced by the Figures, the
precession of the orbit and that of the stream are very similar. Thus
the amount of precession in the debris plane along the stream is
equivalent to the amount of precession of the orbital plane in
time. As a result, we can replace the time along the orbit in
\eqref{eq:precession_rate} with the angle along the stream divided by
the orbital frequency to get
\eq{ \Delta \phi &\approx \omega_{\rm precess} \frac{ \Delta \theta}{\Omega} , \nln &=  \frac{1-q^{-2}}{2}  \cos \psi_0 \Delta \theta . \label{eq:differential_precession}}
Therefore, the flattening and the direction of the symmetry axis of
the potential can both be gleaned from the measurement of the
precession of the stream. As discussed in
\Secref{sec:orbit_comparison}, whether the potential is prolate or
oblate can be determined by comparing the sense of the precession to
the direction of the orbital motion. The amount of nutation expected
along an orbit or a stream can be estimated similarly. From
\eqref{eq:nutation_freq}, we see that in a time the stars propagate
along the stream by $\Delta \theta$, the stream will have undergone a
fraction $\Delta \theta/(q \pi)$ of a nutation. Following the argument
around \eqref{eq:nutation_amp}, during each nutation period, the
inclination angle swings back and forth by $\delta \psi_{\rm nut}$,
moving by $2\delta \psi_{\rm nut}$. Thus, moving along the stream by
$\Delta \theta$ gives a nutation of
\eq{ \Delta \psi & \approx \delta\psi_{\rm nut} \frac{2 \Delta \theta}{q \pi} , \nln &= \frac{1-q^{-2}}{q \pi} \cos\psi_0 \sin\psi_0 \Delta \theta. \label{eq:differential_nutation} }

\begin{figure}
\centering
\includegraphics[width=0.5\textwidth]{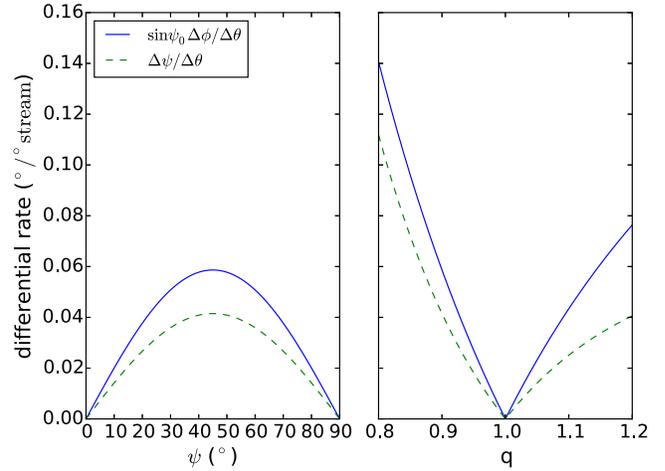}
\caption{Differential precession and nutation rate, i.e. the shift of
  the angular momentum vector in degrees per degree of stream covered,
  in an axisymmetric potential. We show the expected amount of
  precession and nutation along an orbit, or equivalently, a
  stream. The expressions for these rates come from
  \protect\eqref{eq:differential_precession} and
  \protect\eqref{eq:differential_nutation} respectively. The solid
  blue line shows the differential precession rate and the dashed
  green line shows the differential nutation rate.  We have included a
  factor of $\sin \psi_0$ in the differential precession rate to give
  the true angle by which the pole would move on the sky. We see that
  the expected amount of differential precession and nutation are
  similar so any observed change in the stream plane will have
  contributions from both precession and nutation. For the plot on the
  left we take $q=0.9$ and for the plot on the right we take
  $\psi_0=45^\circ$. }
\label{fig:differential_precession_and_nutation}
\end{figure}

\Figref{fig:differential_precession_and_nutation} compares predicted
rates of differential precession and nutation as a function of the
inclination angle (left) and the flattening (right). From inspection
of the Figure, it is obvious that the rate of differential precession
is normally higher than the rate of differential nutation, but not
overwhelmingly so. We stress that this result is derived for near
circular orbits, and caution that as we saw in
\Figref{fig:phi_angle_var}, the nutation is more complicated for
eccentric orbits and appears to have a larger amplitude, quite
probably leading to a larger differential nutation.

\subsubsection{Stream width}

The evolution of the stream width can be summarized as
follows. Streams ought to spread out in flattened potentials; the
cross-section increases the fastest for streams which are farthest
from the symmetry plane. As a result, progenitors on polar orbits
should have the widest streams as illustrated in
\Figref{fig:stream_width_angle_var}. While the derivation has focused
on the evolution of the stream width in time, we can also think about
the evolution of the width along the stream. Since stars near the
progenitor have had less time to precess in the non-spherical
potential, the stream width should increase as we move away from the
progenitor. The exact change in the stream width as we move along the
stream is complicated by the varying angular distance between two
nearby orbits, which will cause the diameter of the debris bundle to
pulsate.

The connection between the stream width and the orbital inclination
angle could provide an independent constraint on the orientation of
the symmetry axes of the Milky Way. In practical terms, if widths of
streams observed in various orientations can be measured and if a
systematic trend in width as a function of orientation can be
detected, the orientation and flattening of the halo can be
determined. Furthermore, as demonstrated in \Secref{sec:triaxial}, our
model also works for streams on short and long axis loops in a
triaxial potential. Thus, if the Milky Way potential is triaxial, we
should expect the narrowest streams to be those with planes aligned
with the short and long axis.

\begin{figure}
\centering
\includegraphics[width=0.5\textwidth]{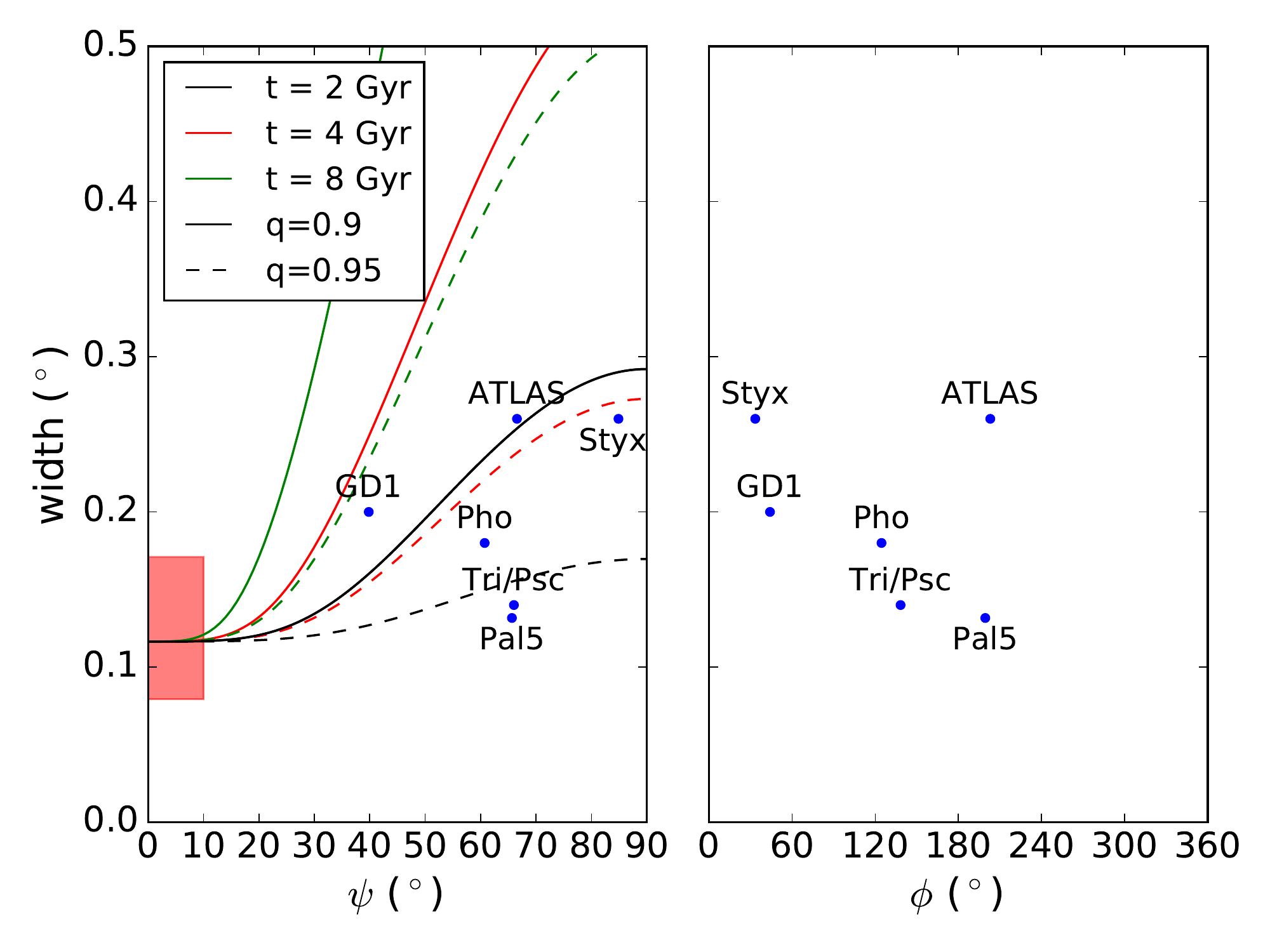}
\caption{Streams widths for seven observed cold (all with presumed
  globular cluster origin) streams as a function of the orientation of
  their plane. The left and right panels respectively show the stream
  width as a function of the polar and azimuthal angle. Small blue
  filled circles and the associated names correspond to the observed
  widths. The solid curves show the expected width from
  \protect\eqref{eq:char_width} for streams with three different ages
  in a potential with $q=0.9$. The dashed curves show the same
  quantities except for potential with $q=0.95$. These curves are
  produced assuming a progenitor with a mass of $10^{4.5}
  M_\odot$. The light red rectangle in the lower left shows the range
  in expected widths for progenitors with a mass of $10^4
  M_\odot$-$10^5 M_\odot$. The bunching of the polar angles of
  observed streams around high inclination values is caused by the
  footprint shapes of the imaging surveys in which they are detected. Note
  that on average, all streams appear sufficiently narrow to exclude
  an effective Galactic potential with $q=0.9$ (but see text and
  Figure~\ref{fig:stream_width_variation} for discussion). Curiously,
  the two streams which are (marginally) the widest (ATLAS and Styx)
  are also the closest to polar orientation. }
\label{fig:stream_widths_observed}
\end{figure}

With these ideas in mind, \Figref{fig:stream_widths_observed} presents
the distribution of observed widths for seven candidate globular
cluster streams in the Milky Way. The seven streams depicted are the
Pal 5 stream \citep {pal5disc}, GD-1 stream \citep{gd1disc}, Styx stream \citep{g_acheron_styx_cocytos_lethe},
Triangulum/Pisces stream \citep[][]{bonacadisc, tripsc_charlesmartin,
  tripsc_pandas}, ATLAS stream \citep{atlasdisc}, and Phoenix stream
\citep{phoenix_disc}. For GD-1 and ATLAS streams, we use the great
circle pole coordinates and the widths reported by
\citet[][]{koposov_et_al_2010} and \citet[][]{atlasdisc}. The widths
of the Phoenix and the Pal 5 streams are taken from
\citet{phoenix_disc} and \citet{ibata_pal5} respectively. For the
remaining streams, we calculate the coordinates of the best-fit
(heliocentric) great circles and the apparent widths ourselves. Given
the great circle pole and the stream's heliocentric distance, we fit
planes to the observable sections of the stream debris to determine
the Galactocentric stream pole angles, $\psi$ and $\phi$. Also shown
are the the stream width model expectations - dictated by
\eqref{eq:char_width} - for various debris ages (in other words,
different time since stripping) which have been evolved in potentials
with $q=0.9$ and $q=0.95$.

A quick glance at \Figref{fig:stream_widths_observed} reveals that the
observed streams tend to congregate towards high inclination angles,
i.e. those with $\psi > 50^{\circ}$. Most likely, this bias is caused
by the footprint shapes of the three imaging surveys the stream data
come from: all three, SDSS, VST ATLAS and DES observed regions of the
sky predominantly above $|b|=30^{\circ}$. Most importantly, however,
all of these streams appear to be remarkably narrow with
$w\approx0.2^{\circ}$. Preference for such low cross-section values is
striking even if by design only cold streams were selected for this
plot. Note that the excluded streams are at least an order of
magnitude broader than those shown. For example, Orphan stream is
$\sim1^{\circ}$ across \citep[see e.g.][]{belokurov2007}, and the
widths of the Sagittarius and Cetus streams are closer to
$\sim10^{\circ}$ \citep[see][]{koposov2012}. Therefore, taken at face
value, clustering of the observed globular cluster streams around such
low width values might imply that the effective flattening of the
Galactic potential (in the volume probed by the streams) is at least
$q\sim0.95$. Reassuringly, the streams which are (marginally) the
widest - ATLAS and Styx - are close to polar orientation, in agreement
with our model, in the assumption that the axis of symmetry in the
Galactic is indeed perpendicular to the disk.

However, interpreting these trends might not be as straightforward as
it seems. First, any potential inference involving the stream widths
relies on the assumption of the progenitor's structural
parameters. The theoretical curves displayed correspond to a single
progenitor with $M=10^{4.5} M_\odot$. In the bottom left we include a
red rectangle which shows the range of initial widths for progenitors
with masses between $10^4 M_\odot$-$10^5 M_\odot$. This uncertainty in
the progenitor's properties would broaden each of the curves into a
band. An additional degeneracy is related to the stream's dynamical age
since the width grows in time as discussed in
\Secref{sec:stream_width}. In this light, it is not at all surprising
that Pal 5's stream has the second narrowest width as the currently
detected debris are located quite close to the
progenitor. Furthermore, note that the model curves in
\Figref{fig:stream_widths_observed} have all assumed the same orbit
for the progenitor. However, as \protect\eqref{eq:char_width}
stipulates, the difference in the orbital frequencies needs to be
taken into account before a robust conclusion can be drawn.

In addition, as described in \Secref{sec:stream_width_growth}, the width of the stream within the stream plane can be comparable to the width perpendicular to the plane, depending on the potential. Since the heliocentric observer is not located within the stream plane, some of the measured cross-section on the sky will also come from the debris spread within the stream plane. The exact contribution from each depends on the observer's orientation and distance relative to the stream plane, as well as the potential. Thus, we must be cautious when interpreting these results and further modelling is required to match the observations.

\begin{figure}
\centering
\includegraphics[width=0.5\textwidth]{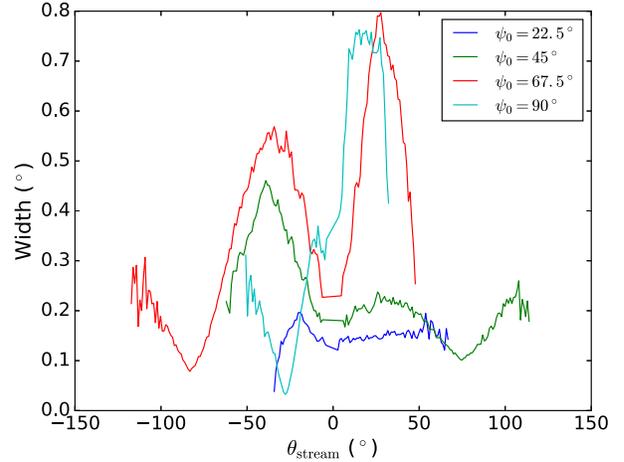}
\caption{Variation of the width along four streams on different
  inclinations in a logarithmic potential with $q=0.9$. The four
  streams are described in \protect\secref{sec:nbody_sims}. The widths
  are computed after 5 Gyr of disruption in $1^\circ$ bins along the stream. We see that the width can
  vary strongly along the stream. Therefore caution must be taken
  when interpreting the widths in
  \protect\figref{fig:stream_widths_observed} since the measurements
  are likely to be biased low, given that the narrow portion of the
  stream are easier to detect given its higher surface
  brightness. Note that stream particles within 2 kpc of the
  progenitor have been masked out.  }
\label{fig:stream_width_variation}
\end{figure}

Furthermore, according to \Secref{sec:stream_width}, the stream width
fluctuates in time (and along the stream) due to the existence of
nodes between the progenitor's plane and the planes of the stream
debris (see \figref{fig:stripping}).
\Figref{fig:stream_width_variation} provides a dramatic illustration
of this effect. Here, the stream width as a function of the angle
along the stream is shown for four simulated streams evolved in a logarithmic
potential with $q=0.9$. As evidenced by the Figure, the extent of the
debris scatter can vary strongly along the stream for highly inclined
orbits. As a result, observations of the stream width may be biased
towards the narrowest parts of the stream which are the easiest to
detect as they are characterized by the highest surface brightness.

These complications show that while the simple model described here is useful for building intuition and understanding the trends, we will need to build a realistic model of the stream which matches its observed properties in order to constrain the shape of the Milky Way \citep[e.g.][]{gibbons_et_al_2014,sanders2014,bovy2014,price_whelan_et_al_rewinder,kuepper_et_al_2015,fardal2015,bowden_gd1}. These models would need to simultaneously match the precession of the stream plane and the stream width to give the most robust constraints.

\subsection{Application to planes of satellites}

Observations of the Milky Way satellites
\citep[e.g.][]{lynden-bell_1976,pawlowski_et_al_2012,pawlowski_et_al_2013},
as well as those around the M31
\citep[e.g.][]{ibata_et_al_2013,conn_et_al_2013}, have uncovered what
appear to be significant anisotropies in the spatial distribution of
the companion dwarf galaxies. These are often interpreted as
preferential planes around which the satellites tend to cluster. If
these satellite planes are indeed genuine, they can perhaps be
explained by group in-fall, the process ubiquitous in Cosmological
structure formation simulations. In that case, the satellite dynamics
and orbital evolution are similar to that of the debris in a tidal
stream since each satellite is on a differently inclined orbit about
the host galaxy. As a result, the results and the intuition developed
here can be applied to interpret the local satellite distribution.
This would suggest that planes aligned with the symmetry axis would be
the longest lived structures and polar planes would be the shortest
lived. Of course, since galactic haloes are likely to be triaxial, the
interpretation is more complicated.

The results of this work mostly agree with those of
\cite{bowden_et_al_2013} who studied the dispersal of satellites in
planar configurations in both axisymmetric and triaxial potentials. In
triaxial potentials, they found that planes of satellites would not
significantly thicken if the normal of the plane was aligned with the
long or short axis, agreeing with the results here. For axisymmetric
potentials they argued that both equatorial and polar orbits will
produce thin disks, in contradiction to the results of this work where
we find that as the misalignment with the symmetry axis increases, the
stream growth rate increases (see \eqref{eq:char_width} and
\Figref{fig:stream_width_angle_var}). One possible reason for this
discrepancy is that \cite{bowden_et_al_2013} only considered one orbit
which was polar. This orbit was in a prolate potential with $q=1.2$ in
the inner region. The other orbits in axisymmetric systems were all
evaluated in oblate potentials with $q=0.8$. From
\eqref{eq:char_width} we see that the stream width goes like
$|q^{-2}-1|$ so we would expect the oblate case in
\cite{bowden_et_al_2013} to broaden twice as fast as their prolate
case.

\subsection{Comparison with previous work}

The framework for the stream width evolution developed in this work
can be used to understand previous results on tidal debris dispersal
as well as those for the precession of
streams. \cite{ibata_et_al_2001,helmi_2004,johnston_et_al_2005} ran
N-body simulations imitating the stream produced by the Sagittarius
dwarf galaxy disruption and found that flattened potentials produce
wider streams. \cite{johnston_et_al_2005} also studied the plane of
the debris and found the same precession and nutation phenomena
discussed in this work. On the observational side, \cite{majewski_et_al_2003} and \cite{belokurov_sgr_precess} found that the pole the Sagittarius stream changes along the stream, likely due to the plane precession described here.

\cite{penarrubia_et_al_2006} studied how streams from tidally
disrupting dwarf galaxies precess in axisymmetric potentials. They
found that the precession rate increased for flatter potentials, as
found here. Similar to \cite{johnston_et_al_2005}, they found that the
trailing tail of a stream precesses slower than the leading
tail. Using the picture in this work, this difference may be due to
the difference in the orbital frequency of the leading and trailing
debris. Since the leading tail has a higher orbital frequency, it can
be expected to precess faster in the host potential. In addition, if
the potential has a flattening which depends on radius, then the
leading and trailing arms can precess at different rates due to the
differences in the Galactic radial range they explore.

\section{Conclusions} \label{sec:conclusion}

We develop a model for the behavior of the orbital angular momentum in
axisymmetric potentials. It has long been known that orbital planes
precess about the potential's symmetry axis while also undergoing a
nutation in the perpendicular direction. Here, we quantify the details of
this precession and nutation in two different limits: one where we
consider a perturbation of a circular orbit in the equatorial plane of
a potential with an arbitrary flattening, and one where the average
torque is computed for a circular orbit with an arbitrary inclination
in a potential with little flattening. This analysis shows that the
precession rate is highest for equatorial orbits and decreases as we
move towards a polar orbit, see i.e. \eqref{eq:precession_rate}. We
also derive expressions for the nutation rate
\eqref{eq:nutation_freq}, as well as the nutation amplitude,
\eqref{eq:nutation_amp}. These analytic results are compared against
numerical integration of both near-circular and eccentric orbits in
\Secref{sec:orbit_comparison} and a good match is found for a large
range of flattening values and orbital inclinations.

We use the results described above to construct a framework for the
evolution of the stream debris plane in axisymmetric potentials. In
the flattened logarithmic potential considered as an example, streams
are shown to precess and nutate like the progenitor's orbit, see
\Figref{fig:orbit_stream_plane} for details. This means that
observations of the twists in the stream plane in the Milky Way can be
used to constrain the flattening and the orientation of the Galactic
potential. Indeed such a motion of the stream angular momentum vector
has already been observed in the Sagittarius stream
\cite{johnston_et_al_2005,belokurov_sgr_precess}. Note, however, that
a disruption of a stellar disk in the Sagittarius dwarf could yield an
apparent stream ``precession'' even in a spherical potential
\citep[][]{gibbons_disk}. Observations of the debris plane twisting in
other streams, especially those far from the Milky Way disk (and those
for which we are certain there was no disky component in the
progenitor), will be invaluable in shedding light onto the properties
of the Milky Way dark matter halo.

The crux of the analysis is the prescription for the stream width
growth in axisymmetric potentials. The idea behind our model is
illustrated in \Figref{fig:stripping} and is based on the differential
plane precession experienced by the debris. The stripped stars are
ejected with slightly different inclination angles relative to the
progenitor. If the potential was spherical, these stars would just
oscillate about the progenitor's orbital plane, resulting in a near
constant stream width. However, in an axisymmetric potential these
differently inclined orbits will precess at different rates, causing
the stream to broaden. For the first time, we derive an expression for
the characteristic stream width in \eqref{eq:char_width}. We
demonstrate that the stream width grows faster as we tilt the stream
plane away from the symmetry axis, with progenitors on polar orbits
producing the broadest streams. We compare our analytic results
against streams in N-body simulations in \Secref{sec:nbody_sims} and
find a good agreement. This stream model can also be used to gain
insight into the dependence of stream widths on the progenitor's
galactocentric radius. We show that the initial spread of the debris
is a weak function of radius in a spherical potential (see
eq. \ref{eq:width_r_dep}), so the stream's physical width is expected
to be proportional to its Galactocentric distance.

While the initial dispersion of the inclination angles of the stripped
stars is almost independent of the progenitor's Galactocentric radius,
the rate of subsequent growth of the debris dispersal is shown to drop
with distance. This is because the stream angular width is
proportional to the characteristic angular frequency of the debris -
this would imply that the trailing tails scatter more slowly than the
leading ones. Most importantly, the amount of accumulated debris
dispersal is a strong and non-linear function of the distance along
the stream. First, recently stripped stars have had less time to move
away from each other due differential plane precession and thus tidal
tails should be narrower near the progenitor. Second, the shape of the
debris profile oscillates along the tidal tail as the angular
separation between the progenitor's orbital plane and the planes of
individual stars changes periodically as explained in Figure
\ref{fig:stripping} and illustrated in Figure
\ref{fig:stream_width_variation}. As evidenced by the latter Figure,
this effect is exacerbated for streams on highly inclined orbits (see
also Figures \ref{fig:stream_width_angle_var} and
\ref{fig:stream_width_q_var}), where the stream width is shown to
experience dramatic transformations. Obviously, such a stream width
evolution implies drastic changes in their surface brightness. This,
in turn, might imply that the current sample of tidal tail detections
in the Galaxy is biased towards the highest surface brightness
portions of otherwise much longer and wider streams.

In \Secref{sec:triaxial}, we extend our analysis to triaxial
potentials and consider specifically short and long axis loop
orbits. These two orbital classes resemble those in axisymmetric
potentials so we might expect that the results derived above would
hold. Indeed, a quantitative comparison of the precession rate as a
function of the inclination from the short and long axis in a triaxial
logarithmic potential presented in \Figref{fig:triax_q_09509} shows a
remarkably good agreement. Additionally, in this potential,
experiments with several simulated streams confirm that the stream width
grows faster as the progenitor's orbit tilts away from the short and long
axis, as expected from the axisymmetric case. Thus, for a restricted
range of orbital classes, the intuition developed in axisymmetric
potentials can be carried over to a triaxial case.

With the stream angular momentum model in place, we discuss a range of the
observable consequences possible. We give expressions for the amount
of precession and nutation along a given section of stream in
\eqref{eq:differential_precession} and
\eqref{eq:differential_nutation} respectively. We find that the amount
of precession and nutation along a stream is comparable, so one must
be cautious when trying to infer the orientation of the
potential. Furthermore, the observed stream widths of seven tidal
tails in the Milky Way halo are shown as a function of the orbital
plane orientation in \Figref{fig:stream_widths_observed}.  These
measurements can be compared to the model under the assumption
that the potential is flattened in the direction normal to the Milky
Way disk. Interestingly, the data shows that the two (marginally)
widest streams (ATLAS and Styx) are also fairly close to
polar. Curiously, it appears difficult to reconcile the apparent
thinness of the detected streams with a flattened potential with
$q=0.9$. Of course, rather than a rigorous analysis, this is merely an
illustration of the application of the model. Evidently, the picture
of debris dispersal based on differential plane precession can also be
applied to the ``planes'' of satellites observed around the Milky Way
and M31. We would expect planes which are aligned with the symmetry
axis of an axisymmetric potential, or the short/long axis of a
triaxial potential, to survive the longest.

In the near future, Gaia and LSST will deliver the data necessary
to take advantage of the concepts introduced here. The sky will be
thoroughly combed for the less obvious stellar streams missed so far
due to their lower surface brightness. Our intuition suggests that
these are expected, both as the result of the disruption of satellites
with higher mass and/or with earlier accretion times. Moreover, even
the streams detected so far might only be short high surface
brightness portions of much longer structures. Be it with true
parallaxes, or ``photometric parallaxes'', good distances are eagerly
expected to complement the existent accurate stream track astrometry
and thus measure the twists of the stream debris planes and their widths. By combining these observations with the intuition developed in this work and with advances in stream modelling \citep[e.g.][]{gibbons_et_al_2014,sanders2014,bovy2014,price_whelan_et_al_rewinder,kuepper_et_al_2015,fardal2015,bowden_gd1} there is hope to finally reveal the shape of the Galactic dark matter halo.

\section*{Acknowledgments}
We thank the Streams group at Cambridge for valuable discussions. We also thank the referee, Kathryn Johnston, for a thoughtful report which improved the discussion in the paper.The research leading to these
results has received funding from the European Research Council under
the European Union's Seventh Framework Programme (FP/2007-2013)/ERC
Grant Agreement no. 308024.

Funding for SDSS-III has been provided by the Alfred P. Sloan Foundation, the Participating Institutions, the National Science Foundation, and the U.S. Department of Energy Office of Science. The SDSS-III web site is \href{http://www.sdss3.org/}{http://www.sdss3.org/}.

SDSS-III is managed by the Astrophysical Research Consortium for the Participating Institutions of the SDSS-III Collaboration including the University of Arizona, the Brazilian Participation Group, Brookhaven National Laboratory, Carnegie Mellon University, University of Florida, the French Participation Group, the German Participation Group, Harvard University, the Instituto de Astrofisica de Canarias, the Michigan State/Notre Dame/JINA Participation Group, Johns Hopkins University, Lawrence Berkeley National Laboratory, Max Planck Institute for Astrophysics, Max Planck Institute for Extraterrestrial Physics, New Mexico State University, New York University, Ohio State University, Pennsylvania State University, University of Portsmouth, Princeton University, the Spanish Participation Group, University of Tokyo, University of Utah, Vanderbilt University, University of Virginia, University of Washington, and Yale University.

\bibliographystyle{mn2e_long}
\bibliography{citations_pp}

\appendix

\section{General Expressions} \label{sec:general_expressions}

For completeness, we also give the precession and nutation rates, as well as the stream growth rates in more general potentials. Instead of the potential expansion in \eqref{eq:potential_q_expansion}, we now consider a potential of the form
\eq{ \Phi(\mathbf{r}) = \Phi(r_0) + f(r_0) \frac{z^2}{r_0} , }
where $r_0 = \sqrt{x^2+y^2+z^2}$. Comparing this with \eqref{eq:potential_q_expansion}, we see that we have an effective $q$ of
\eq{ q_{\rm eff} = \left( 1 + \frac{2 f(r_0)}{\Phi'(r_0)} \right)^{-\frac{1}{2}} . }
We can then plug this $q_{\rm eff}$ back into our expressions for the precession rate \eqref{eq:precession_rate}, nutation rate \eqref{eq:nutation_freq}, nutation amplitude \eqref{eq:nutation_amp}, and characteristic stream width \eqref{eq:char_width} to get
\eq{ \omega_{\rm nutate} &= 2 \Omega \sqrt{  1 + \frac{2 f(r_0)}{\Phi'(r_0)} } , \nln \omega_{\rm precess} &= - \frac{f(r_0)}{r_0 \Omega} \cos\psi_0 , \nln \delta \psi_{\rm nut} &= \frac{f(r_0)}{\Phi'(r_0)} \cos\psi_0 \sin\psi_0 , \nln w &=  \frac{\Delta \psi}{\sqrt{2}} \sqrt{1 + \left( \frac{f(r_0) \sin^2 \psi_0}{r_0 \Omega} t \right)^2}, }
where $\Phi'(r_0) = r_0 \Omega^2$. In this form, the precession and fanning rates can be directly computed for potentials of this form used in the literature, e.g. \cite{bowden_et_al_2013}.

\end{document}